\documentclass[journal,twoside,web]{IEEEtran}

\usepackage[utf8]{inputenc}
\usepackage{color,soul,xcolor}
\usepackage[T1]{fontenc}
\usepackage{multicol}
\usepackage{bm}
\usepackage{multirow}
\usepackage{textcomp}
\usepackage{mathrsfs}
\usepackage{multirow}
\usepackage{graphicx}
\usepackage{epstopdf}
\usepackage{amssymb}
\usepackage{amsmath}
\usepackage{epstopdf}
\usepackage{psfrag}
\usepackage{enumerate}
\usepackage{graphicx}
\usepackage{bigdelim}
\usepackage{amssymb,amsmath,color,psfrag,cuted}
\usepackage{enumerate}
\usepackage{url}
\usepackage{psfrag}
\usepackage{upgreek}
\usepackage{algorithmic,algorithm}
\usepackage{booktabs}
\usepackage{mathrsfs}
\usepackage{array}
\usepackage{multicol}
\usepackage[tight,footnotesize]{subfigure}
\usepackage{multirow}
\usepackage{bm}
\usepackage{soul}
\usepackage{savesym}
\savesymbol{AND}

\usepackage{amsthm}
\usepackage[tight,footnotesize]{subfigure}
\usepackage{cite}
\usepackage{enumerate}
\usepackage{mathtools}
\usepackage{mathabx}
\usepackage{algorithmic,algorithm}

\theoremstyle{plain}

\newtheorem{theorem}{Theorem}
\theoremstyle{definition}

\newtheorem{definition}{Definition}
\newtheorem{lemma}{Lemma}
\newtheorem{corollary}{Corollary}

\newtheorem{remark}{Remark}
\newtheorem{example}{Example}

\makeatletter
\newcommand{\vast}{\bBigg@{3.2}}
\newcommand{\Vast}{\bBigg@{5.5}}

\makeatother

\def\BibTeX{{\rm B\kern-.05em{\sc i\kern-.025em b}\kern-.08em
		T\kern-.1667em\lower.7ex\hbox{E}\kern-.125emX}}

\usepackage{etoolbox}
\makeatletter 
\pretocmd\@bibitem{\color{black}\csname keycolor#1\endcsname}{}{\fail}
\newcommand\citecolor[1]{\@namedef{keycolor#1}{\color{blue}}}
\makeatother
 
\begin{document}

\title{Robust Fixed-Order Controller Design for Uncertain Systems with Generalized Common Lyapunov Strictly Positive Realness Characterization}

\author{
	Jun Ma,  
	Haiyue Zhu, 
	Xiaocong Li, 
	Wenxin Wang,\\
	Clarence W. de Silva, \IEEEmembership{Life Fellow,~IEEE,}
	and Tong Heng Lee	
\thanks{Jun Ma is with the Robotics and Autonomous Systems Thrust, The Hong Kong University of Science and Technology (Guangzhou), Guangzhou, China, also with the Department of Electronic and Computer Engineering, The Hong Kong University of Science and Technology, Hong Kong SAR, China, and also with the HKUST Shenzhen-Hong Kong Collaborative Innovation Research Institute, Futian, Shenzhen, China (e-mail: jun.ma@ust.hk).}
	\thanks{Haiyue Zhu is with the Singapore Institute of Manufacturing Technology, A*STAR, Singapore 138634 (e-mail: zhu\_haiyue@simtech.a-star.edu.sg).}
		\thanks{Xiaocong Li is with the John A. Paulson School of Engineering and Applied Sciences, Harvard University, Cambridge, MA 02138 USA (e-mail: xiaocongli@seas.harvard.edu)}
	\thanks{Wenxin Wang and Tong Heng Lee are with the Department of Electrical and Computer Engineering, National University of Singapore, Singapore 117583 (e-mail: wenxin.wang@u.nus.edu, eleleeth@nus.edu.sg).}
	\thanks{Clarence W. de Silva is with the Department of Mechanical Engineering, The University
		of British Columbia, Vancouver, BC, Canada V6T 1Z4 (e-mail: desilva@mech.ubc.ca).}
	\thanks{This work has been submitted to the IEEE for possible publication. Copyright may be transferred without notice, after which this version may no longer be accessible.}
	 }

\maketitle

\begin{abstract}

	This paper investigates the design of a robust fixed-order controller for single-input-single-output (SISO) polytopic systems with interval uncertainties, with the aim that the closed-loop stability is appropriately ensured and the performance specifications on sensitivity shaping are conformed in a specific finite frequency range. Utilizing the notion of generalized common Lyapunov strictly positive realness (CL-SPRness), the equivalence between strictly positive realness (SPRness) and strictly bounded realness (SBRness) is established; and then the specifications on robust stability and performance are transformed into the SPRness of newly constructed systems and further characterized in the framework of linear matrix inequality (LMI) conditions. The proposed methodology avoids the tedious yet mandatory evaluations of the specifications on all vertices of the uncertain polytopic system in an explicit form. Instead, solving five LMIs exclusively suffices for ensuring the robust stability and performance regardless of the number of vertices, and thus the typically heavy computational burden is considerably alleviated. It is also noteworthy that the proposed methodology additionally provides the necessary and sufficient conditions for this robust controller design with the consideration of a prescribed finite frequency range, and therefore significantly less conservatism is attained in the system performance.
	
\end{abstract}


\section{Introduction}

The research in robust fixed-order controller design continues to attract substantial efforts from control engineers and researchers. It is motivated by the real-time system implementation with a high sampling rate required in many real-world applications, where computational efficiency is crucially important under these circumstances. In the literature, it is noted that such conditions render great difficulties to derive a numerically efficient method, as the stability domain is non-convex for polynomials with an order higher than two~\cite{ackermann2012robust}. Several approaches have been proposed for such fixed-order controller design, such as convex approximation~\cite{karimi2007robust,khatibi2008fixed}, iterative heuristic optimization~\cite{iwasaki1999dual}, etc. It is also notable that the problem is usually formulated as Bilinear Matrix Inequality (BMI) conditions~\cite{safonov1994control,chiu2016method}, and thus it suffers from NP-hardness~\cite{mattei2000sufficient,zhou2020affine}. Besides, another reason that leads to a high computational burden is the inevitable existence of parametric uncertainties. In the existing literature, several approaches are developed to address the robust control problems using Linear Matrix Inequalities (LMIs). In~\cite{gasmi2020robust}, a robust sliding window observer-based controller is proposed, which synthesizes the LMIs with more general and less restrictive conditions. In~\cite{sever2021lmi}, the solution to the guaranteed cost control problem using derivative feedback in the reciprocal state space form is proposed, with sufficient conditions on robust stability and performance derived in terms of LMIs. For these existing robust control approaches, it requires the evaluation of stability and performance over the uncertain domain; and for a polytopic system, each vertex is required to be taken into account~\cite{geromel1994decentralized,geromel1996convex}. Indeed the number of these vertices grows exponentially with the number of parametric uncertainties~\cite{ma2019parameter,ma2021optimal,ma2022symmetric}, and thus the computation is rather costly. The robust stability is appropriately studied based on the techniques such as $\mu$-synthesis~\cite{zhou1998essentials} and small-gain technique~\cite{li2021small}. Generally, the robust stability of a polytopic system is either analyzed in the state space with the notion of quadratic stability, or in the polynomial form by the generalized Kharitonov theorem. However, with these approaches, a non-convex optimization problem is typically an outcome of the formulation. In~\cite{feron1996analysis}, a less conservative approach utilizing a parameter-dependent Lyapunov function is presented, wherein a sufficient condition of stability is given and further highlighted by several LMI conditions. The nonlinear decoupling
methods presented in~\cite{chang2013new,chang2015robust} are seminal works to address non-convex matrix inequality conditions. Substantial works have also been reported on the formulation of LMI conditions arising from the notion of positive realness (PRness) with the development of the Kalman-Yakubovich-Popov (KYP) lemma (also known as the positive real lemma)~\cite{rantzer1994convex,tanaka1999new,hencey2007kyp,khatibi2010hh,salehi2021new}. Notably, PRness is an important property which leads to a wide diversity of analytical developments such as in stability analysis, dissipativity, and passivity. Particularly, it establishes an equivalence between the conditions in the frequency domain for a system to be positive real (PR), an input-output relationship of the system in the time domain, and conditions on the matrices describing the state-space representation of the system. Some related works reveal that the vertices to be checked in robust control design involving interval matrix uncertainties could be reduced~\cite{alamo2008new}. However, if the interval matrix uncertainties appear in the LMI condition in an affine manner, there are still a large number of LMIs to solve. Therefore, the computational burden becomes one of the major challenges in the robust fixed-order controller design.

Another challenge in the robust fixed-order controller design problem is the conservatism arising from a specific viewpoint of the frequency range. In most of the existing approaches, the system performance is attempted over a full frequency range. Nevertheless, in many real-world situations, it is a common situation that only a specific finite frequency range is more realistic with pertinent interest~\cite{li2013heuristic}. For example, as noted in~\cite{iwasaki2003dynamical}, only a feasible frequency range in sensitivity shaping is needed to be considered; as the performance in extreme frequency ranges (either too high or too low) is not the primary objective in a controller design problem. Consequently, those designs from the perspective of the full frequency range then suffer considerably from conservatism. To achieve the sensitivity shaping specifications in a more practically applicable frequency range, weighting functions can be designed and integrated with the control strategies appropriately~\cite{bevrani2015robust} on the basis of the KYP lemma. However, the system order is typically increased significantly, which inevitably causes excessive numerical computational efforts. In this regard, it is of interest to note that the generalized Kalman-Yakubovich-Popov (GKYP) lemma is developed to cater to the system performance in a finite frequency range~\cite{iwasaki2005generalized}, which serves as an extension of the KYP lemma. The GKYP lemma establishes the equivalence between a frequency domain inequality for a transfer function and an LMI associated with its state-space realization. It is thus worthwhile to highlight that while the KYP lemma accounts for the infinite frequency range, the generalized counterpart is capable of limiting the frequency range under consideration to be finite. However, it needs to be noted that the use of the GKYP lemma for sensitivity shaping in a specific finite frequency range only admits a sufficient but not necessary condition for the controller design~\cite{ma2019robust}, due to the use of the convex separation lemma; and it also introduces hyper-parameters that are required to be manually defined before the optimization.

In this work, an effective approach is generalized from~\cite{ma2019robust} to reduce the computational burden and conservatism in fixed-order controller design for uncertain single-input-single-output (SISO) systems, which considers robust stabilization and robust performance under a prescribed finite frequency range. To study the robust stabilization characteristics, the stabilization problem of the resulting closed-loop system is transformed into the SPRness of a newly constructed system. Furthermore, with the notion of generalized common Lyapunov strictly positive realness (CL-SPRness), the equivalence between strictly positive realness (SPRness) and strictly bounded realness (SBRness) is established, whereby the robust performance specifications in terms of sensitivity shaping are also transformed into the SPRness of several newly constructed systems within a stated restricted finite frequency range. Then, necessary and sufficient conditions are presented. Additionally, these SPRness conditions are appropriately constructed with the formulation of several LMIs. Particularly, this approach avoids the otherwise mandatory evaluations of the system stability and performance on all vertices of the polytopic system, instead only one single set of LMIs is needed to solve. Also, the proposed approach allows for more flexibility in addressing finite frequency ranges in design specifications. To sum up, the proposed approach addresses all the commonly encountered challenges as previously highlighted (such as the computational burden and conservatism arising in the traditional robust controller design process). Specifically, it furnishes the design of a robust fixed-order controller for a polytopic system with interval uncertainties, with the aim that the closed-loop stability is appropriately ensured and the performance specifications on sensitivity shaping are conformed in a specific finite frequency range.

The remainder of this paper is organized as follows.
In Section II, the problem statement is given to describe the plant and the fixed-order controller in the presence of interval matrix uncertainties, as well as the design specifications including the robust stability and robust performance in terms of sensitivity shaping requirements. Then, in Section III, the robust stability specification is characterized and expressed by an LMI condition. Next, further developments on robust performance specifications in terms of sensitivity shaping in a finite frequency range are presented in Section IV, and LMI conditions are formulated to construct the robust performance criteria. In Section V, an illustrative example is given with simulation results, and thus the effectiveness of the proposed methodology is demonstrated. Finally, conclusions are drawn in Section VI.

\textbf{Notations:} the symbol $\bm A^{T}$ represents the transpose of the matrix $\bm A$. $\bm I$ represents the identity matrix with appropriate dimensions. $\textup{diag}\{a_1, a_2, \cdots, a_n\}$ represents the diagonal matrix with numbers $a_1, a_2,  \cdots, a_n$ as diagonal entries. $\mathbb{R}$ denotes the set of real matrices. $f \ast g$ means the convolution operation of two functions. $\sigma_\textup{M}(\cdot)$ returns the maximum singular value. The operator $\otimes$ represents the Kronecker product.

\section{Problem Statement}
This section presents the problem statement, which can also be found in~\cite{ma2019robust}. A Single-Input-Single-Output (SISO) plant is represented by an $n$th-order transfer function in continuous time:
\begin{equation}\label{PlantModel1}
\begin{aligned}
P(s)=\frac{b_{1}s^{n-1}+\cdots+b_{n}}{s^{n}+a_{1}s^{n-1}+\cdots+a_{n}},
\end{aligned}
\end{equation}
where $a_{i}$ and $b_{i}$ are uncertain parameters with $a_{i}\in[a_{i}^{l},a_{i}^{u}]$ and $b_{i}\in[b_{i}^{l},b_{i}^{u}]$, $i=1,2,\cdots,n$. Here, we define the medians of these uncertain parameters as $a_{i}^{c}=(a_{i}^{l}+a_{i}^{u})/2$ and $b_{i}^{c}=(b_{i}^{l}+b_{i}^{u})/2$. Also, we define the deviations of these uncertain parameters as $a_{i}^{d}=(a_{i}^{u}-a_{i}^{l})/2$ and $b_{i}^{d}=(b_{i}^{u}-b_{i}^{l})/2$. Thus, these uncertain parameters can be represented in terms of their medians and  deviations, where $a_{i} = a_{i}^{c}+ a_{i}^d \delta_{ai}$, $b_{i} = b_{i}^{c}+ b_{i}^d \delta_{bi}$, $\delta_{ai}\in [-1,1]$ and $\delta_{bi}\in [-1,1]$ are named as standard interval variables. Furthermore, define $\bm\Delta_{\bm a}=\textup{diag}\{\delta_{a1}, \delta_{a2}, \cdots, \delta_{an}\}$ and $\bm\Delta_{\bm b}=\textup{diag}\{\delta_{b1}, \delta_{b2}, \cdots, \delta_{bn}\}$, and then \eqref{PlantModel1} can be equivalently expressed as
\begin{equation}\label{PlantModel2}
\begin{aligned}
P(s)=\frac{(\bm{b^{c}}+[0\quad \bm{b_{d}}\bm{\Delta_{b}}])\bm{s_{n}}^{T}}{(\bm{a^{c}}+[0\quad \bm{a_{d}}\bm{\Delta_{a}}])\bm{s_{n}}^{T}},
\end{aligned}
\end{equation}
with
$\bm{a^{c}}=[1\quad a_{1}^{c}\quad a_{2}^{c}\quad \cdots\quad a_{n}^{c}]$, $\bm{b^{c}}=[0\quad b_{1}^{c}\quad b_{2}^{c}\quad \cdots$ $\quad b_{n}^{c}]$, $\bm{a_{d}}=[a_{1}^{d}\quad a_{2}^{d}\quad \cdots\quad a_{n}^{d}]$, $\bm{b_{d}}=[b_{1}^{d}\quad b_{2}^{d}\quad \cdots\quad b_{n}^{d}]$, $\bm{s_{n}}=[s^{n}\quad s^{n-1}\quad \cdots\ s\quad 1]$. For the sake of simplicity, we define $\bm{a}=\bm{a^{c}}+[\bm{0}\quad \bm{a_{d}}\bm{\Delta_{a}}]$ and $\bm{b}=\bm{b^{c}}+[\bm{0}\quad \bm{b_{d}}\bm{\Delta_{b}}]$.

In this work, an $m$th-order controller is given by
\begin{equation}\label{Controller1}
\begin{aligned}
K(s)=\frac{y_{0}s^{m}+y_{1}s^{m-1}+\cdots+y_{m}}{s^{m}+x_{1}s^{m-1}+\cdots+x_{m}},
\end{aligned}
\end{equation}
where $y_0, y_1, \cdots, y_m$ are coefficients in the numerator polynomial, and  $x_1, \cdots, x_m$ are coefficients in the denominator polynomial.
Also, \eqref{Controller1} can be expressed by
$
K(s)={\bm{y}\bm{s_{m}}^{T}}/{\bm{x}\bm{s_{m}}^{T}},
$
with
$\bm{x}=[1\quad x_{1}\quad x_{2}\quad \cdots\quad x_{m}]$,
$\bm{y}=[y_{0}\quad y_{1}\quad y_{2}\quad \cdots$ $\quad y_{m}]$,
$\bm{s_{m}}=[s^{m}\quad s^{m-1}\quad \cdots \quad s\quad 1]$.

For the closed-loop system, the sensitivity and complementary sensitivity transfer functions $S(s)$ and $T(s)$ are given by
$S(s)={S_{num}(s)}/{S_{den}(s)}$, $T(s)={T_{num}(s)}/{T_{den}(s)},
$
respectively, where
$S_{num}(s) = (\bm{a}\ast\bm{x})\bm{s_{m+n}}^{T}$,
$T_{num}(s) = (\bm{b}\ast\bm{y})\bm{s_{m+n}}^{T}$,
$S_{den}(s)=T_{den}(s) = (\bm{a}\ast\bm{x}+\bm{b}\ast\bm{y})\bm{s_{m+n}}^{T}$,
$
\bm{s_{m+n}}=[s^{m+n}\quad s^{m+n-1}\quad \cdots \quad s\quad 1].
$
Equivalently, we have
\begin{equation}
\begin{aligned}
S_{num}(s) &= (\bm{a^{c}}\ast\bm{x})\bm{s_{m+n}}^{T}+((\bm{a_{d}}\bm{\Delta_{a}})\ast\bm{x})\bm{s_{m+n-1}}^{T},\\
T_{num}(s) &= (\bm{b^{c}}\ast\bm{y})\bm{s_{m+n}}^{T}+((\bm{b_{d}}\bm{\Delta_{b}})\ast\bm{y})\bm{s_{m+n-1}}^{T},\\
S_{den}(s)&=T_{den}(s)=(\bm{a^{c}}\ast\bm{x}+\bm{b^{c}}\ast\bm{y})\bm{s_{m+n}}^{T}\\&\qquad+((\bm{a_{d}}\bm{\Delta_{a}})\ast\bm{x}+(\bm{b_{d}}\bm{\Delta_{b}})\ast\bm{y})\bm{s_{m+n-1}}^{T},
\end{aligned}
\end{equation}
with $\bm{s_{m+n-1}}=[s^{m+n-1}\quad s^{m+n-2}\quad \cdots\quad s\quad 1].$

As a common practice, robust stability and robust performance are primary objectives for control of uncertain systems. It is worthwhile to mention that the robust performance specifications of the closed-loop system can be characterized by sensitivity shaping i.e., $\big| S(j\omega)\big| < \rho_s$, $\forall\omega \in \Omega_s$ and $\big| T(j\omega)\big| < \rho_t$, $\forall\omega \in \Omega_t$, where $\Omega_s$, $\Omega_t$ represent the sets of specific finite frequency range for sensitivity and complementary sensitivity functions, respectively.

\section{Characterization of robust stability specification}
At the beginning, Lemma~\ref{theorem:1} is given to be used in the sequel; and subsequently, Corollary~\ref{theorem:2} is presented.
\begin{lemma}~\cite{ma2019robust}~\label{theorem:1}
Given matrices $\bm{Q}$, $\bm{H_i}$, $\bm{E_i}$, $i=1,2,\cdots,m$   with appropriate dimensions, $\bm Q$ is symmetric, $\bm{\Delta_i}=\textup{diag}\{\delta_{i1}, \delta_{i2}, \cdots, \delta_{in}\}$ with $\delta_{ij}\in [-1,1]$, $i=1,2,\cdots, m$, $j=1,2,\cdots,n$,
\begin{equation}
\bm Q+
\left[\begin{array}{cccc}
\bm 0 & 	\sum \limits_{i=1}^m \bm{E_i}^T\bm{\Delta_i} \bm{H_i}^T\\
\sum \limits_{i=1}^m \bm{H_i}\bm{\Delta_i} \bm{E_i} & \bm 0
\end{array}\right]< 0 \label{eqn:theorem3:1}
\end{equation}
holds if and only if there exist matrices $\bm{R_i}=\textup{diag}\{\varepsilon_{i1}, \varepsilon_{i2},$ $\cdots, \varepsilon_{in}\}$ with $\varepsilon_{ij}>0$, $i=1,2,\cdots, m$, $j=1,2,\cdots,n$, such that
\begin{equation}
\bm Q+
\left[
\begin{array}{cc}
\sum \limits_{i=1}^m \bm{E_i}^T \bm{R_i}^{-1}	\bm{E_i} & \bm 0\\
\bm 0& \sum \limits_{i=1}^m \bm{H_i}  \bm{R_i} \bm{H_i}^T \end{array}
\right]<  0. \label{eqn:theorem2:2}
\end{equation}
\end{lemma}

\begin{corollary}~\label{theorem:2}
Given matrices $\bm Q$, $\bm{H_1}$, $\bm{H_2}$, $\bm{E_1}$, $\bm{E_2}$ with appropriate dimensions, $\alpha$ and $\beta$ are positive real numbers, $\bm{Q}$ is symmetric, $\bm{\Delta_i}=\textup{diag}\{\delta_{i1}, \delta_{i2}, \cdots, \delta_{in}\}$ with $\delta_{ij}\in [-1,1]$, $i=1,2$, $j=1,2,\cdots,n$,
\begin{equation}
\bm Q+
\left[\begin{array}{cccc}
\bm 0 & 	\begin{split} \alpha \bm{E_1}^T\bm{\Delta_1} \bm{H_1}^T\\+\beta \bm{E_2}^T\bm{\Delta_2} \bm{H_2}^T\end{split}\\
\begin{split} \alpha \bm{H_1}\bm{\Delta_1} \bm{E_1}\\+\beta \bm{H_2}\bm{\Delta_2} \bm{E_2} \end{split} &  \bm{0}
\end{array}\right]<  0 \label{eqn:theorem3:1}
\end{equation}
holds if and only if there exist matrices $\bm{R_i}=\textup{diag}\{\varepsilon_{i1}, \varepsilon_{i2},$ $\cdots, \varepsilon_{in}\}$ with $\varepsilon_{ij}>0$, $i=1,2$, $j=1,2,\cdots,n$, such that
\begin{equation}
\bm Q+
\left[
\begin{array}{cc}
\begin{split} \bm{E_1}^T \bm{R_1}^{-1} \bm{E_1}\\+  \bm{E_2}^T \bm{R_2}^{-1} \bm{E_2} \end{split}& \bm{0}\\
\bm{0}&  \begin{split} \alpha^2 \bm{H_1} \bm{R_1} \bm{H_1}^T\\ +\beta^2 \bm{H_2} \bm{R_2} \bm{H_2}^T\end{split} \end{array}
\right]< {0}. \label{eqn:theorem2:2}
\end{equation}
\end{corollary}
\noindent{\textbf{Proof of Corollary~\ref{theorem:2}:}}
First, we denote $\bm{H_1} = \alpha \bm{\bar H_1}$ and $\bm{H_2} = \beta \bm{\bar H_2}$. Then, from Lemma~\ref{theorem:1}, \begin{equation}
\bm Q+
\left[\begin{array}{cccc}
\bm 0 & 	\begin{split}\alpha \bm{E_1}^T\bm{\Delta_1} \bm{\bar H_1}^T\\ +\beta \bm{E_2}^T\bm{\Delta_2} \bm{\bar H_2}^T\end{split} \\
\begin{split}\alpha \bm{\bar H_1} \bm{\Delta_1 E_1} \\+\beta \bm{\bar H_2 \Delta_2 E_2}\end{split} &\bm 0
\end{array}\right]<  0 \label{eqn:theorem1:11}
\end{equation}
holds if and only if there exist matrices $\bm{R_i}=\textup{diag}\{\varepsilon_{i1}, \varepsilon_{i2},$ $\cdots, \varepsilon_{in}\}$ with $\varepsilon_{ij}>0$, $i=1,2 $, $j=1,2,\cdots,n$, such that
\begin{equation}
\bm Q+
\left[
\begin{array}{cc}
\begin{split} \bm{E_1}^T \bm{R_1}^{-1} \bm{E_1}\\+  \bm{E_2}^T \bm{R_2}^{-1} \bm{E_2} \end{split}& \bm 0\\
\bm 0&  \begin{split} \alpha^2
\bm{\bar H_1 R_1} \bm{\bar H_1}^T\\ +\beta^2 \bm{\bar H_2 R_2} \bm{\bar H_2}^T\end{split} \end{array}
\right]<  0. \label{eqn:theorem2:12}
\end{equation}
Then, with~\eqref{eqn:theorem1:11} and~\eqref{eqn:theorem2:12}, Corollary~\ref{theorem:2} is proved by denoting $\bm H_1=\bm{\bar H_1}$ and $\bm H_2=\bm{\bar H_2}$.
\hfill{\qed}

Note that the following developments in this section restate the characterization of robust stability specification as presented in~\cite{ma2019robust}. 
A transfer function is defined as
\begin{equation}\label{DefGs}
\begin{aligned}
G_{s}(s)=\frac{(\bm{a}\ast\bm{x}+\bm{b}\ast\bm{y})\bm{s_{m+n}}^{T}}{d_{c}(s)}=G_{sn}(s)+G_{su}(s),
\end{aligned}
\end{equation}
where $d_c(s)$ is a user-defined strictly Hurwitz polynomial (all roots have strictly negative real parts), $G_{sn}(s)$ and $G_{su}(s)$ are the nominal and uncertain parts of $G_{s}(s)$, respectively, where
\begin{equation}\label{DefGsnp}
\begin{aligned}
G_{sn}(s)=&\frac{(\bm{a^{c}}\ast\bm{x}+\bm{b^{c}}\ast\bm{y})\bm{s_{m+n}}^{T}}{d_{c}(s)},\\
G_{su}(s)=&\frac{((\bm{a_{d}}\bm{\Delta_{a}})\ast\bm{x}+(\bm{b_{d}}\bm{\Delta_{b}})\ast\bm{y})\bm{s_{m+n-1}}^{T}}{d_{c}(s)}.
\end{aligned}
\end{equation}

Basically, the polynomial $d_c(s)$ characterizes the basic desired performance of the closed-loop system, and the controller design approach to be proposed provides extra freedoms such that the performance can be optimized towards the baseline. One typical selection of $d_c(s)$ is to use the characteristic polynomial that is determined from the closed-loop system under practically feasible poles predefined by the user, whereby the controller design approach ensures optimized performance. Normally, a higher-order $d_c(s)$ yields higher complexity in the LMI, which might aggravate the numerical error in the LMI solver and also lead to infeasible or conservative results. Thus, the order of $d_c(s)$ should be kept as low as possible if the results already meet the specifications. Therefore, the selection of $d_c(s)$ can be determined along these lines.

In this work, all the state-space realizations are in the controllable canonical form. Then, the realization of  $G_{sn}(s)$ is expressed as
\begin{equation}\label{realization1}
\sum\nolimits_{sn} \triangleq
\{\bm{A_{sn}}, \bm{B_{sn}}, \bm{C_{sn}}, \bm{D_{sn}}\},
\end{equation}
where $\bm{A_{sn}}, \bm{B_{sn}}, \bm{C_{sn}}, \bm{D_{sn}}$ are the state matrix, input matrix, output matrix, and feedthrough matrix of the state-space model after realization.
Then, $G_s(s)$ is realized in the following form:
\begin{equation}~\label{eq:realization}
\sum\nolimits_{s}  \triangleq
\{
\bm{A_{sn}}, \bm{B_{sn}},
\bm{C_{sn}}+\bm{a_{d}}\bm{\Delta_{a}}\bm{X}+\bm{b_{d}}\bm{\Delta_{b}}\bm{Y}, \bm{D_{sn}}
\},
\end{equation}
where $\bm X$ and $\bm Y$ are the Toeplitz matrices, with
\begin{equation}\label{RealNominal}
\begin{aligned}
\bm X=&
\left[
\begin{array} {ccccccc}
1       & x_{1}   & \cdots    & x_{m}     &0        &0          &0      \\
0       & 1       & x_{1}     & \cdots    & x_{m}   &0          &0      \\
\vdots  &\vdots   &\vdots     &\ddots     &\vdots   &\vdots     &\vdots\\
0       &0        &0          & 1       & x_{1}     & \cdots    & x_{m}
\end{array}
\right] \in \mathbb{R}^{n\times(n+m)},\\
\bm Y=&
\left[
\begin{array} {ccccccc}
y_{0}   & y_{1}   & \cdots    & y_{m}     &0        &0          &0      \\
0       & y_{0}   & y_{1}     & \cdots    & y_{m}   &0          &0      \\
\vdots  &\vdots   &\vdots     &\ddots     &\vdots   &\vdots     &\vdots\\
0       &0        &0          & y_{0}     & y_{1}   & \cdots    & y_{m}
\end{array}
\right] \in \mathbb{R}^{n\times(n+m)}.\\
\end{aligned}
\end{equation}

Note that in~\eqref{eq:realization}, all the parametric uncertainties are imposed on the output matrix only, and this is one of the key points that only one single set of LMIs suffices to address the parametric uncertainties in the whole uncertain domain.

With the notion of SPRness, the system is asymptotically stable if and only if
$
G_{s}(s) \in \mathcal{S},
$
where  $\mathcal{S}$ denotes the set of strictly positive real (SPR) transfer functions. Then, Theorem~\ref{theorem:stability} is given, which establishes the equivalence between the robust stability specification and an LMI condition.

\begin{theorem}~\label{theorem:stability}\cite{ma2019robust}
The robust stability of the system~\eqref{PlantModel1} in the presence of bounded parametric uncertainties characterized by standard interval variables is guaranteed under the controller~\eqref{Controller1} if and only if there exist a Hermitian matrix $\bm{P_s}>0$, diagonal matrices $\bm{R_{sa}}>0$ and $\bm{R_{sb}}>0$ such that
\begin{equation}\label{LMI1}
\begin{aligned}
\left[
\begin{array} {c|c}
\bm{\Gamma_s}  &
\begin{array} {cc}
\bm{X}^T & \bm{Y}^T \\
\bm 0 & \bm 0
\end{array}  \\ \hline
\begin{array} {cc}
\bm X & \bm 0 \\
\bm Y & \bm 0
\end{array}
&
\begin{array} {cc}
-\bm{R_{sa}} & \bm{0}  \\
\bm 0 &  -\bm{R_{sb}}
\end{array}
\end{array}
\right]
<0
\end{aligned},
\end{equation}
where
\begin{equation}
\begin{aligned}
\bm{\Gamma_s} &=
\left[
\begin{array} {cc}
\bm{A_{sn}}^T \bm{P_s} +\bm{P_s} \bm{A_{sn}} & \bm{P_s B_{sn}}  \\
\bm{B_{sn}}^T \bm{P_s}  & \bm 0
\end{array}
\right]  \\
&-\left[
\begin{array} {cc}
\bm 0 & \bm{C_{sn}}^T  \\
\bm C_{sn} & \bm{D_{sn}}+\bm{D_{sn}}^T-\bm{a_{d}} \bm{R_{sa}} \bm{a_{d}}^T-\bm{b_{d}} \bm{R_{sb}} \bm{b_{d}}^T
\end{array}
\right].
\end{aligned}
\end{equation}
\end{theorem}

\noindent{\textbf{Proof of Theorem~\ref{theorem:stability}:}}
From the KYP lemma~\cite{kottenstette2014relationships}, $G_{s}(s) \in \mathcal{S}$ is satisfied if and only if there exists a Hermitian matrix $\bm{P_{s}}>0$ such that
\begin{equation}\label{KYPPR1}
\begin{aligned}
&\left[
\begin{array} {cc}
\bm{A_{sn}}^T \bm{P_s} +\bm{P_s} \bm{A_{sn}} & \bm{P_s B_{sn}}  \\
\bm{B_{sn}}^T \bm{P_s}  & \bm 0
\end{array}
\right] -
\left[
\begin{array} {cc}
\bm 0 & \bm{C_{sn}}^T \\
\bm{C_{sn}}& \bm{D_{sn}}+\bm{D_{sn}}^{T}
\end{array}
\right] \\
&+\left[
\begin{array} {cc}
\bm 0 &  -\bm{X}^T  \bm{\Delta_{a}}  \bm{a_{d}}^T  -\bm{Y}^T \bm{\Delta_{b}}  \bm{b_{d}}^T \\
-\bm{a_{d}} \bm{\Delta_{a}} \bm{X}-\bm{b_{d}} \bm{\Delta_{b}} \bm{Y} & \bm{0}
\end{array}
\right]<0.
\end{aligned}
\end{equation}

From Lemma~\ref{theorem:1}, \eqref{KYPPR1} holds if and only if there exist positive definite diagonal matrices $\bm{R_{sa}}$ and $\bm{R_{sb}}$ such that
\begin{equation}~\label{eq:oo}
\begin{aligned}
&\left[
\begin{array} {cc}
\bm{A_{sn}}^T \bm{P_s} +\bm{P_s A_{sn}} & \bm{P_s B_{sn}}  \\
\bm{B_{sn}}^T \bm{P_s}  & \bm{0}
\end{array}
\right] -
\left[
\begin{array} {cc}
\bm 0 & \bm{C_{sn}}^T  \\
\bm{C_{sn}} & \bm{D_{sn}}+\bm{D_{sn}}^T
\end{array}
\right]\\
&+\left[
\begin{array} {cc}
\bm{X}^T \bm{R_{sa}}^{-1} \bm{X} +\bm{Y}^T \bm{R_{sb}}^{-1} \bm{Y}   & \bm 0 \\
\bm 0&  \bm{a_{d}} \bm{R_{sa}} \bm{a_{d}}^{T}+ \bm{b_{d}}\bm{R_{sb}}\bm{b_{d}}^{T}
\end{array}
\right]
<0.
\end{aligned}
\end{equation}
Then, \eqref{eq:oo} can be equivalently expressed as
the LMI condition \eqref{LMI1}.    \hfill{\qed}

\section{Characterization of finite frequency robust performance specifications}

To characterize the finite frequency range, $\bm\Phi$ is defined as
$
\bm \Phi=
\left[
\begin{array} {cc}
0 & 1\\
1 & 0
\end{array}
\right]$. Also, as shown in Table 1, $\bm{\Psi}$ is given according to different ranges of $\Omega$. Then, with an appropriate choice of $\bm\Psi$, certain ranges of the frequency variable can be defined. Note that more details on frequency range characterization can be referred in~\cite{iwasaki2005generalized}. In Table I, we define $\omega_c=({\omega_h+\omega_l})/{2}$ for the sake of simplicity.  

\begin{table}[h]\centering\normalsize~\label{tab:freq}
\caption{Related matrices for frequency range characterization}
\begin{tabular}{cccc}
	\toprule
	$\Omega$ & $\omega\in(0,\,\omega_l)$ & $\omega\in(\omega_l,\,\omega_h)$ & $\omega\in(\omega_h,\,+\infty)$ \\
	\midrule
	$\bm\Psi$ &
	$\left[
	\begin{array} {cc}
	-1 & 0\\
	0 &  \omega_l^2
	\end{array}
	\right]$
	&
	$\left[
	\begin{array} {cc}
	-1 & j \omega_c\\
	-j \omega_c & -\omega_l\omega_h
	\end{array}
	\right]$
	&
	$\left[
	\begin{array} {cc}
	1 & 0\\
	0 & -\omega_h^2
	\end{array}
	\right]$
	\\
	\bottomrule
\end{tabular}
\end{table}

For sensitivity shaping, the infinity norm of the sensitivity and complementary functions are bounded by certain values. The following lemma presents the results to convert the SBRness condition on the infinity norm to the SPRness condition.

\begin{lemma}~\label{lemma:ND}
The SBRness property
\begin{equation}\label{FFBR0}
\begin{aligned}
\left\|\frac{N(j\omega)}{D(j\omega)}\right\|_{\infty}<\gamma, \, \forall\omega \in \Omega
\end{aligned}
\end{equation}
holds if and only if
\begin{equation}\label{FFPR}
\begin{aligned}
\textup{Re}\left(\frac{D(j\omega)-\gamma^{-1}N(j\omega)}{D(j\omega)+\gamma^{-1}N(j\omega)}\right)>0, \, \forall\omega \in \Omega.
\end{aligned}
\end{equation}
\end{lemma}

\noindent{\textbf{Proof of Lemma~\ref{lemma:ND}:}}
Necessity: \eqref{FFPR} implies that for all $\omega  \in \Omega$, it gives
\begin{equation}
\begin{aligned}\label{qqq}
\frac{D(j\omega)-\gamma^{-1}N(j\omega)}{D(j\omega)+\gamma^{-1}N(j\omega)}+\frac{D^{*}(j\omega)-\gamma^{-1}N^{*}(j\omega)}{D^{*}(j\omega)+\gamma^{-1}N^{*}(j\omega)}> 0,
\end{aligned}
\end{equation}
which leads to
\begin{equation}\label{eq:ND}
\begin{aligned}
\frac{2 \left[D^{*}(j\omega) D(j\omega)-\gamma^{-2}N^{*}(j\omega) N(j\omega)\right] }{\left[D(j\omega)+\gamma^{-1}N(j\omega)\right]\left[D^{*}(j\omega)+\gamma^{-1}N^{*}(j\omega)\right]}  > 0.
\end{aligned}
\end{equation}
It is straightforward that 
\begin{equation}\begin{aligned}\label{eq:NDDD} D^{*}(j\omega) D(j\omega)-\gamma^{-2}N^{*}(j\omega) N(j\omega) > 0, \, \forall\omega \in \Omega.\end{aligned}
\end{equation}
Then, we have
\begin{equation}\label{}
\begin{aligned}
\frac{N^{*}(j\omega)N(j\omega)}{D^{*}(j\omega)D(j\omega)} <\gamma^{2}, \, \forall\omega \in \Omega.
\end{aligned}
\end{equation}
and it gives the SBRness property~\eqref{FFBR0}.

Sufficiency: The SBRness property~\eqref{FFBR0} yields \eqref{eq:NDDD}. Then, it is straightforward that \eqref{eq:ND} holds, which then gives \eqref{qqq}. Therefore, it implies that the condition \eqref{FFPR} holds. This completes the proof of Lemma~\ref{lemma:ND}.  \hfill{\qed}

Define the transfer functions $G_{p}(s)$ as
\begin{equation}\label{DefS1S2}
\begin{aligned}
G_{p}(s)=&G_{pn}(s)+G_{pu}(s),
\end{aligned}
\end{equation}
where
\begin{equation}\label{DefS1S2np}
\begin{aligned}
G_{pn}(s)=&\frac{(\bm{a^{c}}\ast\bm{x})\bm{s_{m+n}}^{T}}{d_{c}(s)},\\
G_{pu}(s)=&\frac{((\bm{a_{d}}\bm{\Delta_{a}})\ast\bm{x})\bm{s_{m+n-1}}^{T}}{d_{c}(s)}.
\end{aligned}
\end{equation}
Given that $G_{pn}(s)$ is realized in the following form:
\begin{equation}\label{realization2}
\begin{aligned}
\sum \nolimits_{pn}&\triangleq
\{\bm{A_{pn}}, \bm{B_{pn}},	\bm{C_{pn}}, \bm{D_{pn}}\},
\end{aligned}
\end{equation}where $\bm{A_{pn}}, \bm{B_{pn}}, \bm{C_{pn}}, \bm{D_{pn}}$ are the state matrix, input matrix, output matrix, and feedthrough matrix of the state-space model after realization, and then $G_{p}(s)$ is realized in the following form:
\begin{equation}\label{realization22}
\begin{aligned}
\sum \nolimits_{p}&\triangleq
\{
\bm{A_{pn}}, \bm{B_{pn}}, \bm{C_{pn}}+\bm{a_{d}} \bm{\Delta_{a}} \bm{X}, \bm{D_{pn}}
\}.
\end{aligned}
\end{equation}

As $S(j\omega)= G_p(j\omega)/G(j\omega)$, it follows from Lemma~\ref{lemma:ND} that $\big| S(j\omega)\big| < \rho_s$, $\forall \omega\in\Omega_s$  is equivalent to
\begin{equation}
\begin{aligned}
\textup{Re}\left(\frac{G_s(j\omega)-\rho_s^{-1}G_p(j\omega)}{G_s(j\omega)+\rho_s^{-1}G_p(j\omega)}\right)>0, \forall \omega\in\Omega_s.
\end{aligned}
\end{equation}
The definitions of common Lyapunov strictly positive real (CL-SPR) and common Lyapunov stable (CL-stable) are given in~\cite{khatibi2008fixed}. In this work, to address the finite frequency specifications, generalized notions of these definitions are used, which are defined over a finite frequency range. 
\begin{definition}
	Two finite frequency SPR transfer functions $H_1$ and $H_2$ with controllable canonical state space realizations $(\bm{A_1},\bm{B_1},\bm{C_1},\bm{D_1})$ and $(\bm{A_2},\bm{B_2},\bm{C_2},\bm{D_2})$ are called generalized CL-SPR, if both satisfy the inequality of the GKYP lemma under the stated finite frequency range with the same Lyapunov pair $(\bm{P},\bm{Q})$. 
\end{definition}
\begin{definition}
	Consider two monic polynomials $p_1$ and $p_2$ and their state-state matrices $\bm{A_1}$ and $\bm{A_2}$, respectively. Then, $p_1$ and $p_2$ (also $\bm{A_1}$ and $\bm{A_2}$) are called generalized CL-stable, if $\bm{A_1}$ and $\bm{A_2}$ satisfy the following inequalities under the stated finite frequency range with the same Lyapunov pair $(\bm{P},\bm{Q})$:
	\begin{equation}~\label{eq:31}
	\begin{aligned}
	\bm{A_1}^T \bm{P}+\bm{P}\bm{A_1}&+ \Psi_{11} \bm{A_1}^T  \bm{Q} \bm{A_1}+\Psi_{12} \bm{A_1}^T  \bm{Q}\\&+\Psi_{21}\bm{Q}\bm{A_1}+\Psi_{22}\bm{Q}<0,\\
	\bm{A_2}^T \bm{P}+\bm{P}\bm{A_2}&+ \Psi_{11} \bm{A_2}^T  \bm{Q} \bm{A_2}+\Psi_{12} \bm{A_2}^T  \bm{Q}\\&+\Psi_{21}\bm{Q}\bm{A_2}+\Psi_{22}\bm{Q}<0.
	\end{aligned}
	\end{equation}
\end{definition}
Notice that referring to Table 1, $\bm{\Psi}$ is essentially a symmetric matrix under the listed three different ranges of $\Omega$. In~\eqref{eq:31}, $\Psi_{11}$, $\Psi_{12}$, $\Psi_{21}$, and $\Psi_{22}$ represent the elements in the matrix $\bm\Psi$. Take the case when $\omega\in(0,\,\omega_l)$ as an example, we have $\Psi_{11}=-1$, $\Psi_{12}=0$, $\Psi_{21}=0$, and $\Psi_{22}=\omega_l^2$. Thus in this sense, we have 
\begin{eqnarray}
    &\bm{A_1}^T \bm{P}+\bm{P}\bm{A_1}+ \Psi_{11} \bm{A_1}^T  \bm{Q} \bm{A_1}+\Psi_{12} \bm{A_1}^T  \bm{Q}+\Psi_{21}\bm{Q}\bm{A_1}\nonumber\\&+\Psi_{22}\bm{Q}
    = \bm{A_1}^T \bm{P}+\bm{P}\bm{A_1}-\bm{A_1}^T  \bm{Q} \bm{A_1}+\omega_l^2 \bm{Q},
    \end{eqnarray}
which is symmetric. Along this line, for the scenarios when $\omega\in(\omega_l,\,\omega_h)$ and $\omega\in(\omega_h,\,+\infty)$, $\bm{A_1}^T \bm{P}+\bm{P}\bm{A_1}+ \Psi_{11} \bm{A_1}^T \bm{ Q} \bm{A_1}+\Psi_{12} \bm{A_1}^T  \bm{Q}+\Psi_{21}\bm{Q}\bm{A_1}+\Psi_{22}\bm{Q}$ remains symmetric too, and so does $\bm{A_2}^T \bm{P}+\bm{P}\bm{A_2}+ \Psi_{11} \bm{A_2}^T  \bm{Q} \bm{A_2}+\Psi_{12} \bm{A_2}^T  \bm{Q}+\Psi_{21}\bm{Q}\bm{A_2}+\Psi_{22}\bm{Q}$. 

Next, Lemma~\ref{lemma:SPR} is introduced to be used in the sequel, and then Lemma~\ref{theorem:CLSPR} is presented to reveal the relationship between the finite frequency SPRness condition and generalized CL-SPRness.
\begin{lemma}~\cite{khatibi2010hh}\label{lemma:SPR}
	The following statements hold:
	\begin{enumerate}[(a)]	\item A transfer function is SPR if and only if its numerator and denominator are CL-stable.
		\item If two transfer functions are CL-SPR, then all of their numerator and denominator polynomials are CL-stable.
	\end{enumerate}
\end{lemma}

\begin{lemma}~\label{theorem:CLSPR} The finite frequency SPRness condition
\begin{equation}~\label{eqn:posreal}
\begin{aligned}
\textup{Re}\left(\frac{G_s(j\omega)-\rho_s^{-1}G_p(j\omega)}{G_s(j\omega)+\rho_s^{-1}G_p(j\omega)} \right)>0, \forall \omega \in \Omega
\end{aligned}
\end{equation} is satisfied if and only if ${G_s(j\omega)-\rho_s^{-1}G_p(j\omega)}$ and ${G_s(j\omega)+\rho_s^{-1}G_p(j\omega)} $ are generalized CL-SPR.
\end{lemma}	
\noindent{\textbf{Proof of Lemma~\ref{theorem:CLSPR}:}}
Essentially, the claims in Lemma~\ref{lemma:SPR} can be extended to the generalized scenarios, where SPR, CL-stable and CL-SPR are generalized to fit for the case in a finite frequency range. With these supporting results, the proof of Lemma~\ref{theorem:CLSPR} is given.

Necessity: for all $\omega\in \Omega$, if ${G_s(j\omega)-\rho_s^{-1}G_p(j\omega)}$ and ${G_s(j\omega)+\rho_s^{-1}G_p(j\omega)} $ are generalized CL-SPR, from the generalized results of Lemma~\ref{lemma:SPR}, all the numerator and denominator polynomials of ${G_s(j\omega)-\rho_s^{-1}G_p(j\omega)}$ and ${G_s(j\omega)+\rho_s^{-1}G_p(j\omega)}$ are generalized CL-stable.  As ${G_s(j\omega)-\rho_s^{-1}G_p(j\omega)}$ and ${G_s(j\omega)+\rho_s^{-1}G_p(j\omega)}$ have the same denominator polynomial $d_c(s)$, then, in $\frac{G_s(j\omega)-\rho_s^{-1}G_p(j\omega)}{G_s(j\omega)+\rho_s^{-1}G_p(j\omega)} $, $d_c(s)$ is canceled out with only the numerator polynomials of ${G_s(j\omega)-\rho_s^{-1}G_p(j\omega)}$ and ${G_s(j\omega)+\rho_s^{-1}G_p(j\omega)}$ left. Indeed, the numerator polynomials of  ${G_s(j\omega)-\rho_s^{-1}G_p(j\omega)}$ and ${G_s(j\omega)+\rho_s^{-1}G_p(j\omega)}$ are generalized CL-stable. Thus, again from the generalized results of Lemma~\ref{lemma:SPR}, we have \eqref{eqn:posreal}.

Sufficiency: we prove it by contradiction. If \eqref{eqn:posreal} is satisfied, it can be seen that the numerator polynomials of ${G_s(j\omega)-\rho_s^{-1}G_p(j\omega)}$ and ${G_s(j\omega)+\rho_s^{-1}G_p(j\omega)}$ are generalized CL-stable (because their denominator polynomials $d_c(s)$ are canceled out). Suppose that ${G_s(j\omega)-\rho_s^{-1}G_p(j\omega)}$ and ${G_s(j\omega)+\rho_s^{-1}G_p(j\omega)} $ are not generalized CL-SPR, then the numerator polynomial of ${G_s(j\omega)-\rho_s^{-1}G_p(j\omega)}$, the numerator polynomial of ${G_s(j\omega)+\rho_s^{-1}G_p(j\omega)}$, and the denominator polynomial $d_c(j\omega)$ are not generalized CL-stable. Hence, it is impossible that both ${G_s(j\omega)-\rho_s^{-1}G_p(j\omega)}$ and ${G_s(j\omega)+\rho_s^{-1}G_p(j\omega)}$ are finite frequency SPR, which contradicts the fact that $d_c(s)$ is a strictly Hurwitz polynomial. Thus, the sufficiency is proved.
\hfill{\qed}

From Lemma~\ref{theorem:CLSPR}, it can be shown that ${G_s(j\omega)-\rho_s^{-1}G_p(j\omega)}$ and ${G_s(j\omega)+\rho_s^{-1}G_p(j\omega)}$ satisfy the GKYP lemma with the same Lyapunov matrix $\bm P$ and  associated matrix $\bm Q$, which are so-called a Lyapunov pair. It is notable that $\bm{A_{sn}}=\bm{A_{pn}}$, $\bm{B_{sn}}=\bm{B_{pn}}$, $\bm{D_{sn}}=\bm{D_{pn}}$, and $\text{dim}(\bm{C_{sn}})=\text{dim}(\bm{C_{pn}})$. Then, we have
\begin{align}
&\,\quad{G_s(s)+\rho_s^{-1}G_p(s)} \nonumber \\
&=
\left[(\bm{C_{sn}}+\rho_s^{-1}\bm{C_{pn}}+(1+\rho_s^{-1})\bm{a_{d}}\bm{\Delta_{a}}\bm{X}+\bm{b_{d}}\bm{\Delta_{b}}\bm{Y})\right] \nonumber\\
&\,\quad(s\bm I-\bm{A_{sn}})^{-1}\bm{B_{sn}}+(1+\rho_s^{-1}) \bm{D_{sn}}.
\end{align}
Here, we denote $\rho_{s}^+=1+\rho_s^{-1}$ and $\rho_{s}^-=1-\rho_s^{-1}$, then the realizations of ${G_s(s)+\rho_s^{-1}G_p(s)}$ and  ${G_s(s)-\rho_s^{-1}G_p(s)}$ are given by
\begin{equation}
\begin{aligned}
\sum\nolimits_p^{+}  &\triangleq
\{
\bm{A_{sn}}, \bm{B_{sn}},
\bm{C_{sn}}+\rho_s^{-1}\bm{C_{pn}}+\rho_{s}^+\bm{a_{d}}\bm{\Delta_{a}}\bm{X}  \\
&\quad+\bm{b_{d}}\bm{\Delta_{b}}\bm{Y}), \rho_{s}^+\bm{D_{sn}}
\},\\
\sum\nolimits_p^{-}  &\triangleq
\{
\bm{A_{sn}}, \bm{B_{sn}},
\bm{C_{sn}}-\rho_s^{-1}\bm{C_{pn}}+\rho_{s}^-\bm{a_{d}}\bm{\Delta_{a}}\bm{X}  \\
&\quad+\bm{b_{d}}\bm{\Delta_{b}}\bm{Y}), \rho_{s}^- \bm{D_{sn}}
\}.
\end{aligned}
\end{equation}
Based on the above development and analysis, Theorem~\ref{theorem:performance} formulates the robust performance specification on the sensitivity function as LMI conditions.
\begin{theorem}~\label{theorem:performance}
The robust performance specification $\big| S(j\omega)\big| < \rho_s$, $\forall \omega\in\Omega_s$ of the system~\eqref{PlantModel1} in the presence of bounded parametric uncertainties characterized by standard interval variables is guaranteed under the controller~\eqref{Controller1} if and only if there exist Hermitian matrices $\bm{P_p}$ and $\bm{Q_p}>0$, diagonal matrices $\bm{R_{pa}}>0$, $\bm{R_{pb}}>0$, $\bm{R_{pc}}>0$, and $\bm{R_{pd}}>0$ such that 		
\begin{equation}\label{LMI2}
\begin{aligned}
\left[
\begin{array} {c|c}
\bm{\Gamma_{pa}}  &
\begin{array} {cc}
\bm{X}^T & \bm{Y}^T \\
\bm{0} & \bm{0}
\end{array}  \\ \hline
\begin{array} {cc}
\bm{X} & \bm{0} \\
\bm{Y} & \bm{0}
\end{array}
&
\begin{array} {cc}
-\bm{R_{pa}} & \bm{0}  \\
\bm{0} &  -\bm{R_{pb}}
\end{array}
\end{array}
\right]
<0
\end{aligned},
\end{equation}
and
\begin{equation}\label{LMI3}
\begin{aligned}
\left[
\begin{array} {c|c}
\bm{\Gamma_{pb}}  &
\begin{array} {cc}
\bm{X}^T & \bm{Y}^T \\
\bm{0} & \bm{0}
\end{array}  \\ \hline
\begin{array} {cc}
\bm{X} & \bm{0} \\
\bm{Y} & \bm{0}
\end{array}
&
\begin{array} {cc}
-\bm{R_{pc}}  & \bm{0}  \\
\bm{0} &  -\bm{R_{pd}}
\end{array}
\end{array}
\right]
<0
\end{aligned},
\end{equation}
where
\begin{equation}
\begin{aligned}
&\bm{\Gamma_{pa}}  =
\left[
\begin{array} {cc}
\bm{A_{sn}} & \bm{B_{sn}} \\
\bm{I} & \bm{0}
\end{array}
\right]^T
\bm{\Upxi_{p}}
\left[
\begin{array} {cc}
\bm{A_{sn}} & \bm{B_{sn}}  \\
\bm{I} & \bm{0}
\end{array}
\right] \\
&-\left[
\begin{array} {cc}
\bm{0} & \bm{C_{sn}}^T + \rho_s^{-1}\bm{C_{pn}}^T \\
\bm{C_{sn}}+\rho_s^{-1}\bm{C_{pn}} & \begin{split} \rho_s^+(\bm{D_{sn}}+\bm{D_{sn}}^T)&-(\rho_s^+)^2\bm{a_{d}} \bm{R_{pa}} \bm{a_{d}}^T \\&-\bm{b_{d}} \bm{R_{pb}} \bm{b_{d}}^T \end{split}
\end{array}
\right],
\end{aligned}
\end{equation}
\begin{equation}
\begin{aligned}
&\bm{\Gamma_{pb}}  =
\left[
\begin{array} {cc}
\bm{A_{sn}} & \bm{B_{sn}} \\
\bm{I} & \bm{0}
\end{array}
\right]^T
\bm{\Upxi_{p}}
\left[
\begin{array} {cc}
\bm{A_{sn}} & \bm{B_{sn}}  \\
\bm{I} & \bm{0}
\end{array}
\right] \\
&-\left[
\begin{array} {cc}
\bm{0} & \bm{C_{sn}}^T - \rho_s^{-1}\bm{C_{pn}}^T \\
\bm{C_{sn}}-\rho_s^{-1}\bm{C_{pn}} & \begin{split} \rho_s^-(\bm{D_{sn}}+\bm{D_{sn}}^T)&-(\rho_s^-)^2\bm{a_{d}} \bm{R_{pc}} \bm{a_{d}}^T \\&-\bm{b_{d}} \bm{R_{pd}} \bm{b_{d}}^T \end{split}
\end{array}
\right],
\end{aligned}
\end{equation}
$\bm{\Upxi_{p}}=\bm{\Phi_s}\otimes \bm{P_{p}} +\bm{\Psi_s}\otimes \bm{Q_{p}}$, $\bm{\Phi_s}$ and $\bm{\Psi_s}$ are matrices that characterize the frequency range $\Omega_s$.
\end{theorem}	

\noindent{\textbf{Proof of Theorem~\ref{theorem:performance}:}}
As   ${G_s(j\omega)+\rho_s^{-1}G_p(j\omega)}$ satisfies the GKYP lemma, we have
\begin{equation}\label{KYPPR111}
\begin{aligned}
&			\left[
\begin{array} {cc}
\bm{A_{sn}} & \bm{B_{sn}} \\
\bm{I} & \bm{0}
\end{array}
\right]^T
\bm{\Upxi_{p}}
\left[
\begin{array} {cc}
\bm{A_{sn}} & \bm{B_{sn}}  \\
\bm{I} & \bm{0}
\end{array}
\right] \\
&-\left[
\begin{array} {cc}
\bm{0} &  \bm{C_{sn}}^T+\rho_s^{-1}\bm{C_{pn}}^T \\
\bm{C_{sn}}+\rho_s^{-1}\bm{C_{pn}} &  \rho_s^{+}(\bm{D_{sn}}+\bm{D_{sn}}^T)
\end{array}
\right] \\
&+\left[
\begin{array} {cc}
\bm{0} &   -\rho_s^{+}\bm{X}^T  \bm{\Delta_{a}}  \bm{a_{d}}^T  -\bm{Y}^T \bm{\Delta_{b}}  \bm{b_{d}}^T \\
-\rho_s^{+}\bm{a_{d}} \bm{\Delta_{a}} \bm{X}-\bm{b_{d}} \bm{\Delta_{b}} \bm{Y} & \bm{0}
\end{array}
\right] \\
&<{0}.
\end{aligned}
\end{equation}
From Corollary~\ref{theorem:2}, \eqref{KYPPR111} holds if and only if there exist positive definite diagonal matrices $\bm R_{pa}$ and $\bm R_{pb}$ such that
\begin{equation}\label{KYPPR112}
\begin{aligned}
&			\left[
\begin{array} {cc}
\bm{A_{sn}} & \bm{B_{sn}} \\
\bm{I} & \bm{0}
\end{array}
\right]^T
\bm{\Upxi_{p}}
\left[
\begin{array} {cc}
\bm{A_{sn}} & \bm{B_{sn}}  \\
\bm{I} & \bm{0}
\end{array}
\right] \\
&-\left[
\begin{array} {cc}
\bm{0} &  \bm{C_{sn}}^T+\rho_s^{-1}\bm{C_{pn}}^T \\
\bm{C_{sn}}+\rho_s^{-1}\bm{C_{pn}} & \rho_s^{+}(\bm{D_{sn}}+\bm{D_{sn}}^T)
\end{array}
\right] \\
&+\left[
\begin{array} {cc}
\bm{X}^T \bm{R_{pa}}^{-1} {\bm X} +\bm{Y}^T \bm{R_{pb}}^{-1} \bm{Y}   & \bm{0} \\
\bm{0}&  (\rho_s^{+})^2 \bm{a_{d}} \bm{R_{pa}} \bm{a_{d}}^{T}+ \bm{b_{d}}\bm{R_{pb}}\bm{b_{d}}^{T}
\end{array}
\right]\\
&<0.
\end{aligned}
\end{equation}
Then, we have
\begin{equation}\label{KYPPR2}
\begin{aligned}
&			\left[
\begin{array} {cc}
\bm{A_{sn}} & \bm{B_{sn}} \\
\bm{I} & \bm{0}
\end{array}
\right]^T
\bm{\Upxi_{p}}
\left[
\begin{array} {cc}
\bm{A_{sn}} & \bm{B_{sn}}  \\
\bm{I} & \bm{0}
\end{array}
\right]\\
&-
\left[
\begin{array} {cc}
\bm{0} & \bm{C_{sn}}^T+\rho_s^{-1}\bm{C_{pn}}^T   \\
\bm{C_{sn}}+\rho_s^{-1}\bm{C_{pn}}  & \rho_s^{+}(\bm{D_{sn}}+\bm{D_{sn}}^T)
\end{array}
\right]\\
& +(\rho_s^+)^2\left[
\begin{array} {c}
\bm{0} \\
\bm{a_{d}}
\end{array}
\right]\bm{R_{pa}}
\left[
\begin{array} {cc}
\bm{0} &  \bm{a_{d}}^{T}
\end{array}
\right]
+
\left[
\begin{array} {c}
\bm{0} \\
\bm{b_{d}}
\end{array}
\right]\bm{R_{pb}}
\left[
\begin{array} {cc}
\bm{0} &  \bm{b_{d}}^{T}
\end{array}
\right]
\\
&+ \left[
\begin{array} {c}
\bm{X}^T \\
\bm 0
\end{array}
\right] \bm{R_{pa}}^{-1}
\left[
\begin{array} {cc}
\bm{X} & \bm{0}
\end{array}
\right]
+
\left[
\begin{array} {c}
\bm{Y}^T \\
\bm{0}
\end{array}
\right] \bm{R_{pb}}^{-1}
\left[
\begin{array} {cc}
\bm{Y} & \bm{0}
\end{array}
\right]
<0.
\end{aligned}
\end{equation}
\eqref{KYPPR2} can be equivalently expressed 
in the form of \eqref{LMI1}. Similarly, ${G_s(j\omega)-\rho_s^{-1}G_p(j\omega)}$ satisfies the GKYP lemma with the same Lyapunov pair, then it is straightforward to derive \eqref{LMI2}.   \hfill{\qed}

Furthermore, the results on the specification in terms of the sensitivity function can be extended to the specification on the complementary sensitivity function. Thus define transfer functions $G_{q}(s)$  as
\begin{equation}\label{qqq1}
\begin{aligned}
G_{q}(s)=&G_{qn}(s)+G_{qu}(s),
\end{aligned}
\end{equation}
where
\begin{equation}\label{qqq2}
\begin{aligned}
G_{qn}(s)=&\frac{(\bm{b^{c}}\ast\bm{y})\bm{s_{m+n}}^{T}}{d_{c}(s)},\\
G_{qu}(s)=&\frac{((\bm{b_{d}}\bm{\Delta_{b}})\ast\bm{y})\bm{s_{m+n-1}}^{T}}{d_{c}(s)}.
\end{aligned}
\end{equation}
Similarly, $| T(j\omega)|<\rho_t$, $\forall \omega\in\Omega_t$ is satisfied if and only if ${G_s(j\omega)-\rho_t^{-1}G_q(j\omega)}$ and ${G_s(j\omega)+\rho_t^{-1}G_q(j\omega)} $ are generalized CL-SPR.

$G_{qn}(s)$ can be realized in the controllable canonical form as
\begin{equation}\label{realization3}
\begin{aligned}
\sum \nolimits_{qn}&\triangleq
\{\bm{A_{qn}}, \bm{B_{qn}},	\bm{C_{qn}}, \bm{D_{qn}}\},
\end{aligned}
\end{equation} where $\bm{A_{qn}}, \bm{B_{qn}}, \bm{C_{qn}}, \bm{D_{qn}}$ are the state matrix, input matrix, output matrix, and feedthrough matrix of the state-space model after realization. Then, the state-space realization of $G_{q}(s)$ is given by
\begin{equation}\label{realization33}
\sum \nolimits_{q} \triangleq
\{
\bm{A_{qn}}, \bm{B_{qn}}, \bm{C_{qn}}+\bm{b_{d}} \bm{\Delta_{b}} \bm{Y}, \bm{D_{qn}}
\}.
\end{equation}
Remarkably, we have  $\bm{A_{sn}}=\bm{A_{qn}}$, $\bm{B_{sn}}=\bm{B_{qn}}$, $\bm{D_{qn}}=\bm{0}$, and $\text{dim}(\bm{C_{sn}})=\text{dim}(\bm{C_{qn}})$. Denote $\rho_{t}^+=1+\rho_t^{-1}$ and $\rho_{t}^-=1-\rho_t^{-1}$, then the realizations of ${G_s(s)+\rho_t^{-1}G_q(s)}$ and ${G_s(s)-\rho_t^{-1}G_q(s)}$ are given by
\begin{equation}
\begin{aligned}
\sum\nolimits_q^{+}  &\triangleq
\{
\bm{A_{sn}}, \bm{B_{sn}},
\bm{C_{sn}}+\rho_t^{-1}\bm{C_{qn}}+\bm{a_{d}}\bm{\Delta_{a}}\bm{X}  \\
&\quad+\rho_{t}^+\bm{b_{d}}\bm{\Delta_{b}}\bm{Y}),  \bm{D_{sn}}
\},\\
\sum\nolimits_q^{-}  &\triangleq
\{
\bm{A_{sn}}, \bm{B_{sn}},
\bm{C_{sn}}-\rho_t^{-1}\bm{C_{qn}}+\bm{a_{d}}\bm{\Delta_{a}}\bm{X}  \\
&\quad+\rho_{t}^-\bm{b_{d}}\bm{\Delta_{b}}\bm{Y}),   \bm{D_{sn}}
\}.
\end{aligned}
\end{equation}

Next, Theorem~\ref{theorem:performance2} is given, which relates the robust performance specification on complementary sensitivity shaping to LMI conditions.

\begin{theorem}~\label{theorem:performance2}
The robust performance specification $\big| T(j\omega)\big| < \rho_t$, $\forall \omega\in\Omega_t$ of the system~\eqref{PlantModel1} in the presence of bounded parametric uncertainties characterized by standard interval variables is guaranteed under the controller~\eqref{Controller1} if and only if there exist Hermitian matrices $\bm{P_q}$ and $\bm{Q_q}>0$, diagonal matrices $\bm{R_{qa}}>0$, $\bm{R_{qb}}>0$, $\bm{R_{qc}}>0$, and $\bm{R_{qd}}>0$ such that 		
\begin{equation}\label{LMI4}
\begin{aligned}
\left[
\begin{array} {c|c}
\bm{\Gamma_{qa}}  &
\begin{array} {cc}
\bm{X}^T & \bm{Y}^T \\
\bm{0} & \bm{0}
\end{array}  \\ \hline
\begin{array} {cc}
\bm{X} & \bm{0} \\
\bm{Y} & \bm{0}
\end{array}
&
\begin{array} {cc}
-\bm{R_{qa}} & \bm{0}  \\
\bm{0} &  -\bm{R_{qb}}
\end{array}
\end{array}
\right]
<0
\end{aligned},
\end{equation}
and
\begin{equation}\label{LMI5}
\begin{aligned}
\left[
\begin{array} {c|c}
\bm{\Gamma_{qb}}  &
\begin{array} {cc}
\bm{X}^T & \bm{Y}^T \\
\bm{0} & \bm{0}
\end{array}  \\ \hline
\begin{array} {cc}
\bm{X} & \bm{0} \\
\bm{Y} & \bm{0}
\end{array}
&
\begin{array} {cc}
- \bm{R_{qc}} & \bm{0}  \\
\bm{0} &  -\bm{R_{qd}}
\end{array}
\end{array}
\right]
<0
\end{aligned},
\end{equation}
where
\begin{equation}
\begin{aligned}
\bm{\Gamma_{qa}} &=
\left[
\begin{array} {cc}
\bm{A_{sn}} & \bm{B_{sn}} \\
\bm{I} & \bm{0}
\end{array}
\right]^T
\bm{\Upxi_{q}}
\left[
\begin{array} {cc}
\bm{A_{sn}} & \bm{B_{sn}}  \\
\bm{I} & \bm{0}
\end{array}
\right]  \\
&-\left[
\begin{array} {cc}
\bm{0} & \bm{C_{sn}}^T + \rho_t^{-1}\bm{C_{qn}}^T \\
\bm{C_{sn}}+\rho_t^{-1}\bm{C_{qn}} & \begin{split} \bm{D_{sn}}+\bm{D_{sn}}^T&-\bm{a_{d}} \bm{R_{qa}} \bm{a_{d}}^T \\&-(\rho_t^+)^2\bm{b_{d}} \bm{R_{qb}} \bm{b_{d}}^T \end{split}
\end{array}
\right],
\end{aligned}
\end{equation}
\begin{equation}
\begin{aligned}
\bm{\Gamma_{qb}} &=
\left[
\begin{array} {cc}
\bm{A_{sn}} & \bm{B_{sn}} \\
\bm{I} & \bm{0}
\end{array}
\right]^T
\bm{\Upxi_{q}}
\left[
\begin{array} {cc}
\bm{A_{sn}} & \bm{B_{sn}}  \\
\bm{I} & \bm{0}
\end{array}
\right]  \\
&-\left[
\begin{array} {cc}
\bm{0} & \bm{C_{sn}}^T - \rho_t^{-1}\bm{C_{qn}}^T \\
\bm{C_{sn}}-\rho_t^{-1}\bm{C_{qn}} & \begin{split}  \bm{D_{sn}}+\bm{D_{sn}}^T&-\bm{a_{d}} \bm{R_{qc}} \bm{a_{d}}^T \\&-(\rho_t^-)^2\bm{b_{d}} \bm{R_{qd}} \bm{b_{d}}^T \end{split}
\end{array}
\right],
\end{aligned}
\end{equation}
$\bm{\Upxi_{q}}=\bm{\Phi_t}\otimes \bm{P_{q}} +\bm{\Psi_t}\otimes \bm{Q_{q}}$, $\bm{\Phi_t}$ and $\bm{\Psi_t}$ are matrices that characterize the frequency range $\Omega_t$.
\end{theorem}	

\noindent{\textbf{Proof of Theorem~\ref{theorem:performance2}:}}	
The proof of Theorem~\ref{theorem:performance2} proceeds along with the same procedures as Theorem~\ref{theorem:performance}.\hfill{\qed}

The proposed algorithm for fixed-order controller design toward uncertain systems is summarized as Algorithm 1. With the results presented in Theorems~\ref{theorem:stability}-\ref{theorem:performance2}, the robust stability and performance is characterized by LMIs \eqref{LMI1}, \eqref{LMI2}, \eqref{LMI3}, \eqref{LMI4}, and \eqref{LMI5}, and thus it facilitates the effective use of several numerical algorithms and solvers.

\begin{algorithm}\label{algo}
	\centering
	\caption{Algorithm for Fixed-Order Controller Design Toward Uncertain Systems}
	\begin{algorithmic}[1]\label{algorithm}
		\REQUIRE
		 $n$th-order SISO plant $P(s)$ in \eqref{PlantModel1} with  $a_{i}\in[a_{i}^{l},a_{i}^{u}]$, $b_{i}\in[b_{i}^{l},b_{i}^{u}]$;
robust performance requirements $\big| S(j\omega)\big| < \rho_s$, $\forall\omega \in \Omega_s$,  $\big| T(j\omega)\big| < \rho_t$, $\forall\omega \in \Omega_t$.\\ 
 
		\STATE Rewrite the plant model $P(s)$ in \eqref{PlantModel2}.
		\STATE Construct  Toeplitz matrices $\bm{X}$ and $\bm{Y}$.
		\STATE Determine $\rho_{s}^+=1+\rho_s^{-1}$, $\rho_{s}^-=1-\rho_s^{-1}$, $\rho_{t}^+=1+\rho_t^{-1}$, $\rho_{t}^-=1-\rho_t^{-1}$.
		\STATE Determine $\bm{\Phi_s}$, $\bm{\Psi_s}$, $\bm{\Phi_t}$, $\bm{\Psi_t}$ according to Table I.
		\STATE Choose a strictly Hurwitz polynomial $d_c(s)$.
		\STATE Define transfer functions $G_s(s)$, $G_{sn}(s)$, $G_p(s)$, $G_{pn}(s)$, $G_q(s)$, $G_{qn}(s)$ by \eqref{DefGs}, \eqref{DefGsnp}, \eqref{DefS1S2}, \eqref{DefS1S2np}, \eqref{qqq1}, \eqref{qqq2}.
		\STATE Realize the transfer functions $G_{sn}(s)$, $G_s(s)$,   $G_{pn}(s)$, $G_p(s)$, $G_{qn}(s)$, $G_q(s)$  by \eqref{realization1}, \eqref{eq:realization}, \eqref{realization2}, \eqref{realization22}, \eqref{realization3}, \eqref{realization33}.
	
		\STATE Solve LMIs \eqref{LMI1}, \eqref{LMI2}, \eqref{LMI3}, \eqref{LMI4}, \eqref{LMI5}.
 \RETURN $m$th-order controller $K(s)$ in \eqref{Controller1}.
	\end{algorithmic}
\end{algorithm}

\begin{remark}
		For an uncertain polytopic system, all extreme systems need to be considered in a typical loop shaping problem. Generally, when the number of parametric uncertainties increases, the number of extreme systems increases exponentially, so a large number of LMIs remain to solve in a typical formulation. On the contrary, our proposed approach synthesizes the loop shaping problem using five LMIs only (one for the requirement of robust stability, two for the requirement of sensitivity function, and two for the requirement of complementary sensitivity function), without leading to significant computational burden due to the exponential increase in the number of extreme systems. Thus, our approach avoids the explicit evaluation of the specifications on all the extreme systems; and in this manner, simpler computational processes are resulted from the design of a fixed-order robust controller in the presence of parametric uncertainties.
	\end{remark}
		
\section{Illustrative example}

To illustrate the effectiveness of the proposed methodology, the following example is presented, which is adapted from~\cite{keel1999robust}. In this problem, the LMIs \eqref{LMI1}, \eqref{LMI2}, \eqref{LMI3}, \eqref{LMI4}, and \eqref{LMI5} are solved by the YALMIP Toolbox in MATLAB, and the simulations are conducted with a laptop processor Intel(R) Core(TM) CPU i7-5600U@2.60GHz.
\begin{example}
An uncertain second-order plant is given by
\begin{equation}
P(s)=\frac{ b_1s +b_2  }{s^2+a_1s+a_2},
\end{equation}
with $a_1\in [0.5,1]$, $a_2 \in [-1,1]$, $b_1 \in [0.5,1]$, $b_2 \in [1,1.5]$. It is aimed to design a second-order controller
\begin{equation}
K(s)=\frac{y_0s^2+y_1s+y_2}{s^2+x_1s+x_2},
\end{equation}
such that the robust stability is ensured in the presence of parametric uncertainties, the magnitude of the sensitivity function $\big| S(j\omega)\big|<-3$ \textup{dB} in the frequency range $\omega \in ($\textup{0.01} \textup{rad/s}, \textup{0.1} \textup{rad/s}$)$, and the magnitude of the complementary sensitivity function $\big| T(j\omega)\big|<-3$ \textup{dB} in the frequency range $\omega \in ($\textup{50} \textup{rad/s}, \textup{100} \textup{rad/s}$)$.
\end{example}

\begin{figure}[t]
\centering
\includegraphics[trim=0 0 0 0,width=0.75\columnwidth]{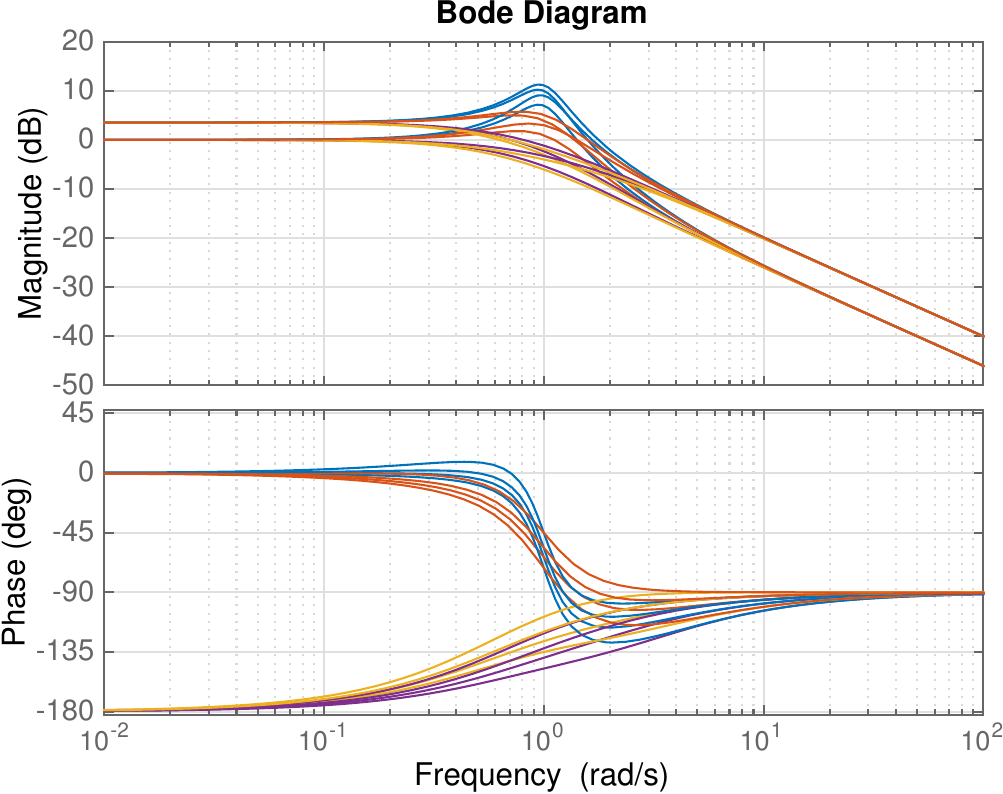}
\caption{Bode diagrams of the plant under parametric uncertainties.}
\label{fig:bodeplant_uncer}
\end{figure}

In view of the parametric uncertainties in the plant, there are 16 polytopic systems, because $a_1$, $a_2$, $b_1$, and $b_2$ are all within the prescribed upper bound and lower bound. To demonstrate the effect of these uncertainties on the system, Fig.~\ref{fig:bodeplant_uncer} shows the bode diagrams of these 16 polytopic systems.

Furthermore, $d_c(s)$ can be chosen by the characteristic polynomial that is determined from the closed-loop system under practically feasible poles, and the order of $d_c(s)$ should be kept as low as possible if the results already meet the specifications. Along these lines, a strictly Hurwitz polynomial is chosen as $d_c(s)= s^4+4.5s^3+6.225s^2+4.525s+1.5$ in this work, which yields the poles at $-0.77$, $-2.77$, and $-0.48\pm0.68i$, such that the performance baseline is governed with this setting. To the industrial preference in terms of practical implementation, $x_2$ is set to be zero, and by solving the LMIs, the fixed-order controller parameters are given by $x_1=0.8213$, $y_0=20.0270$, $y_1=18.3422$, $y_2=18.4318$. In the simulation, random plant parameters are generated within the uncertain bound, and one set of these parameters is given by $\tilde a_1=0.7863$, $\tilde a_2=0.4128$, $\tilde b_1=0.6132$, and $\tilde b_2=1.4309$, which constructs the uncertain plant $\tilde P(s)$. 

\begin{figure}[t]
\centering
\includegraphics[trim=0 0 0 0,width=0.75\columnwidth]{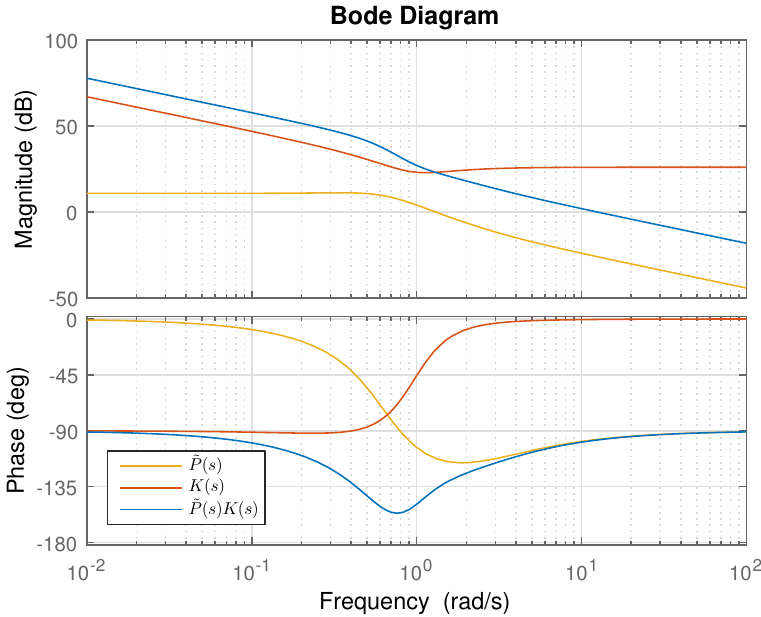}
\caption{Bode diagrams of the plant, the controller, and the open-loop system with $x_2= 0$ using the proposed robust controller synthesis approach}
\label{fig:openloop}
\end{figure}

\begin{figure}[t]
\centering
\includegraphics[trim=0 0 0 0,width=0.75\columnwidth]{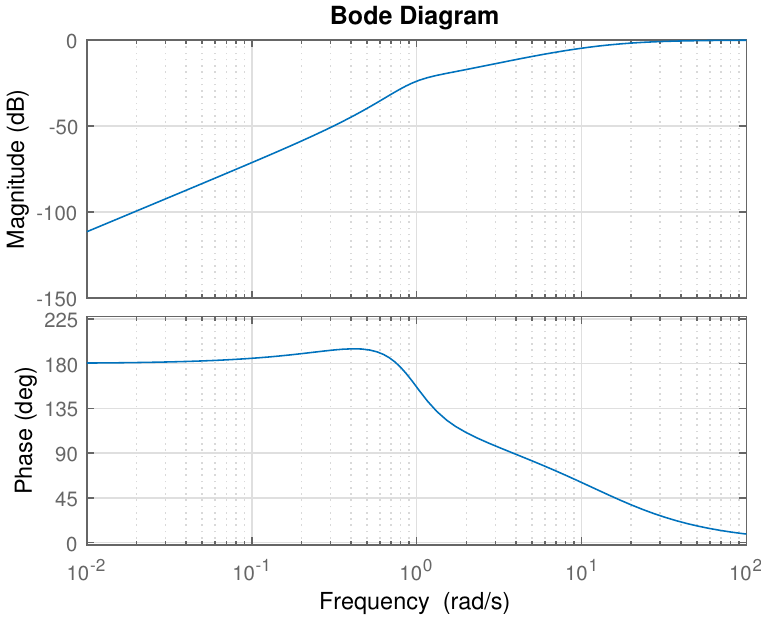}
\caption{Bode diagram of the sensitivity function with $x_2= 0$ using the proposed robust controller synthesis approach}
\label{fig:sensitivity}
\end{figure}

\begin{figure}[t]
\centering
\includegraphics[trim=0 0 0 0,width=0.75\columnwidth]{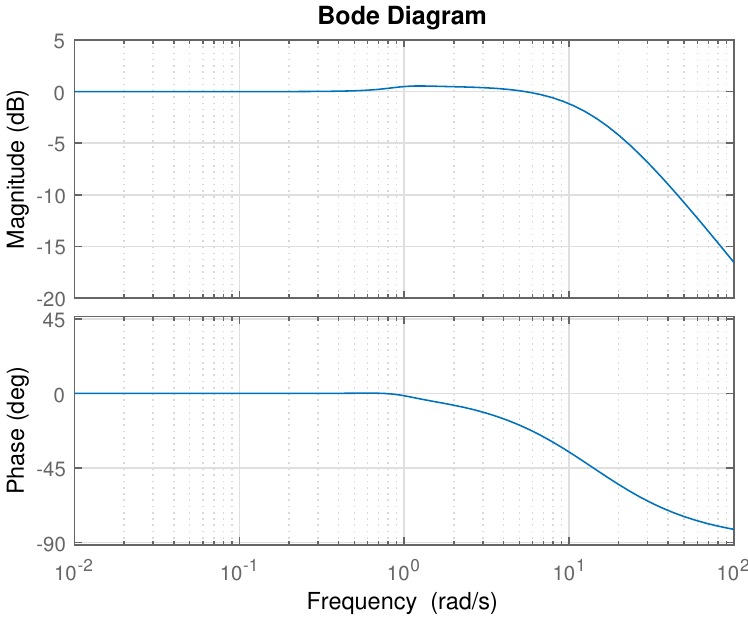}
\caption{Bode diagram of the complementary sensitivity function with $x_2= 0$ using the proposed robust controller synthesis approach}
\label{fig:csensitivity}
\end{figure}

The robust stability can be easily verified against the existence of parametric uncertainties. Furthermore, the bode diagrams of the uncertain plant $\tilde P(s)$, the controller $K(s)$, and the open-loop system $\tilde P(s)K(s)$ are shown in Fig.~\ref{fig:openloop}. Also, Fig.~\ref{fig:sensitivity} and Fig.~\ref{fig:csensitivity} depict the bode diagrams of the sensitivity and complementary sensitivity functions, respectively, and they clearly show that the design specifications on robust performance in terms of the sensitivity and complementary sensitivity functions are met. 

To demonstrate the effectiveness of the proposed development in terms of computational efficiency and system performance, a comparison study is conducted. In the following, our proposed approach in this work is denoted by Method 1. Also, another two approaches based on the GKYP lemma are used for comparative purposes, and the Hurwitz polynomial is kept the same. Note that the approach presented in~\cite{ma2019robust} is denoted by Method 2, which is a more conservative design approach as it provides only a sufficient condition for the performance guarantee. Moreover, the approach presented in~\cite{zhu2019dual} is denoted by Method 3, which gives the fixed-order controller design principle considering all the extreme systems in the uncertain domain without loss of conservatism.  The number of parametric uncertainties in the plant model is 4, and thus the number of LMIs in Method 3 is 80 (16 sets), while both Method 1 and Method 2 give only 5 LMIs (1 set) to solve. 
Generally, it is the typical case that the size of each LMI in these methods is the same, respectively.
By solving these LMIs, the processing time (solvertime) in all these methods is recorded, which is given by 0.7672 s, 1.5872 s, and 256.6223 s, respectively. From the above results, it is clear that our proposed method is slightly more efficient than Method 2, but it shows significant improvement compared with Method 3.

\begin{figure}[t]
	\centering
	\includegraphics[trim=0 0 0 0,width=0.75\columnwidth]{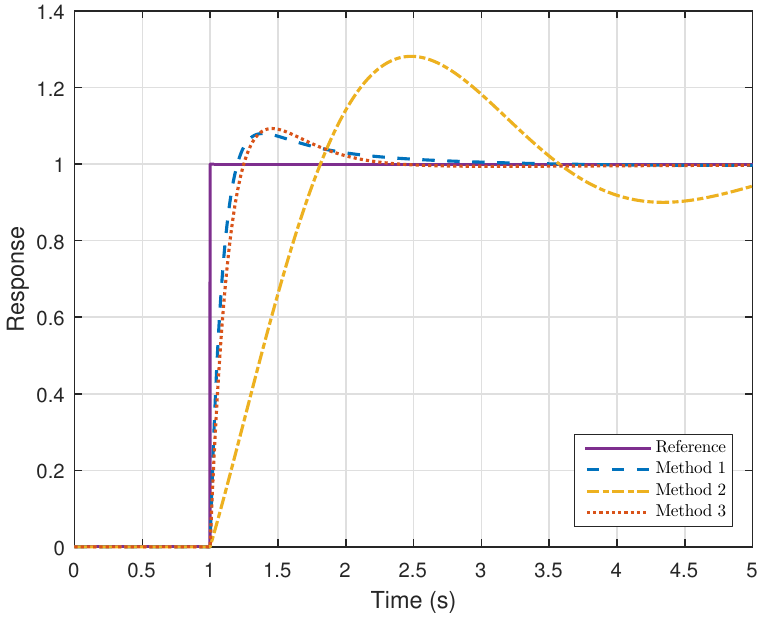}
	\caption{Comparison results of the time response with $x_2= 0$ using Method 1 (the proposed robust controller synthesis approach), Method 2, and Method 3}
	\label{fig:per}
\end{figure}

With Method 2, the controller parameters are given by $x_1=0.7811$, $y_0=1.9825$, $y_1=2.8963$, $y_2=1.1919$; with Method 3, the controller parameters are given by $x_1=0.2514$, $y_0=14.4982$, $y_1=7.1815$, $y_2=1.5031$. Next, a simulation study in the time domain is presented, where a step tracking problem is investigated, and the details are depicted in Fig.~\ref{fig:per}. The values of the Root-Mean-Square Error (RMSE) attained using these three methods are given by 0.0901,    0.2573,  and  0.1035, respectively. The maximum overshoot is given by 7.98\%,    28.14\%,  and  9.28\%, respectively. For the Steady-State Error (SSE), it can be easily calculated, which is given by 0 for all these methods.  It can be clearly observed that Method 1 and Method 3 give rather good system performance, while Method 2 suffers from conservatism. Comparing Method 1 and Method 3, Method 3 is much more time-consuming. Based on the discussions above, our claims and contributions are successfully validated.


\begin{figure}[t]
	\centering
	\includegraphics[trim=0 0 0 0,width=0.75\columnwidth]{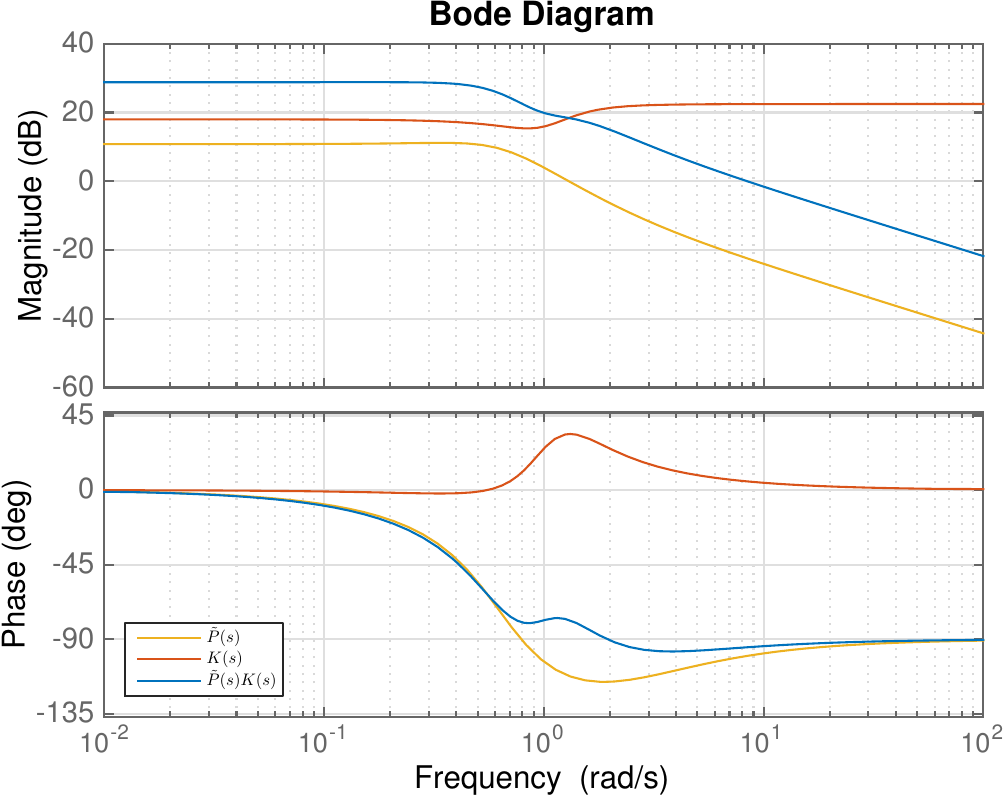}
	\caption{Bode diagrams of the plant, the controller, and the open-loop system with $x_2\neq 0$ using the proposed robust controller synthesis approach}
	\label{fig:openloop_x2nonzero}
\end{figure}

\begin{figure}[t]
	\centering
	\includegraphics[trim=0 0 0 0,width=0.75\columnwidth]{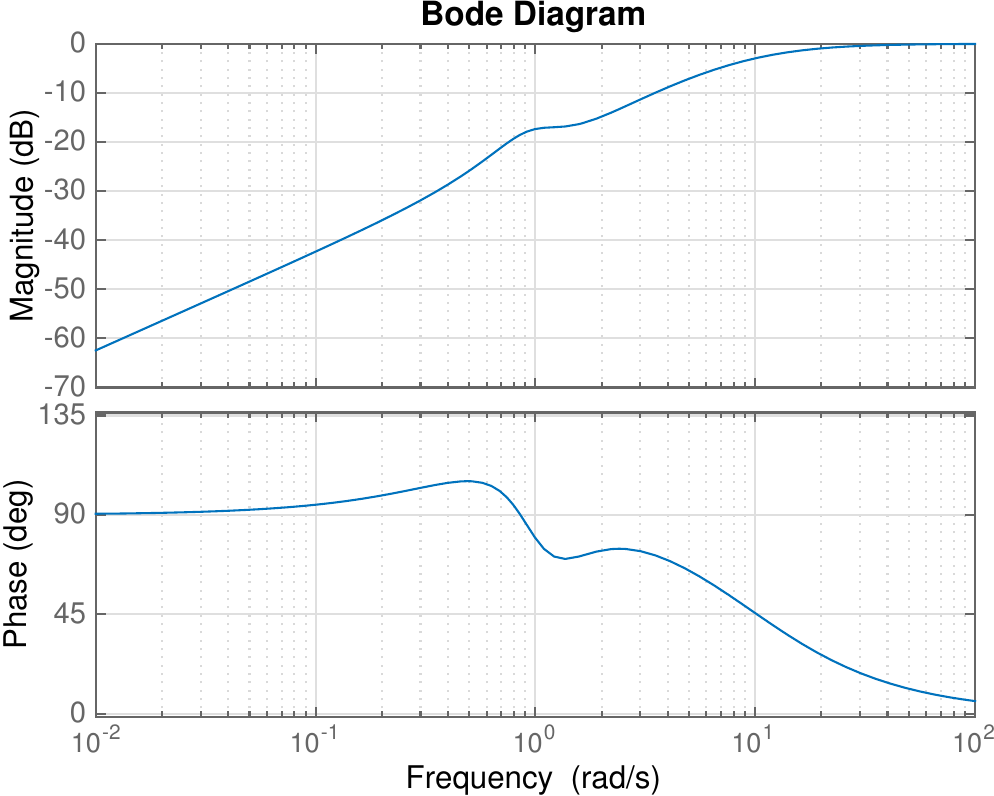}
	\caption{Bode diagram of the sensitivity function with $x_2\neq 0$ using the proposed robust controller synthesis approach}
	\label{fig:sensitivity_x2nonzero}
\end{figure}

\begin{figure}[t]
	\centering
	\includegraphics[trim=0 0 0 0,width=0.75\columnwidth]{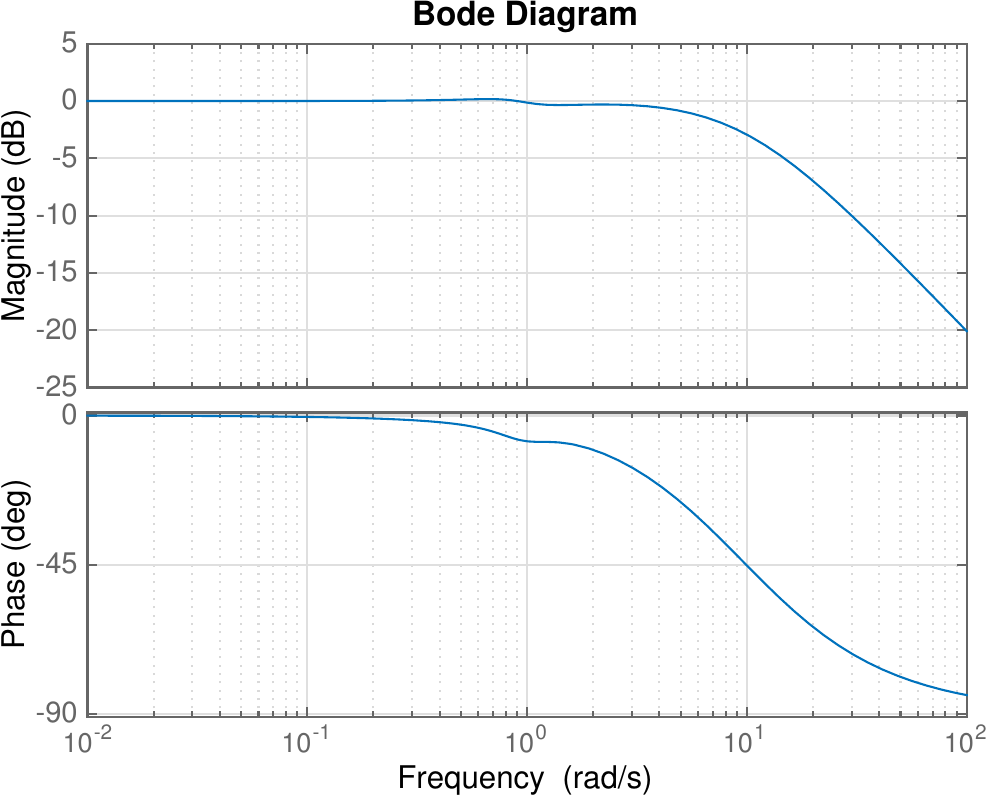}
	\caption{Bode diagram of the complementary sensitivity function with $x_2\neq 0$ using the proposed robust controller synthesis approach}
	\label{fig:csensitivity_x2nonzero}
\end{figure}

\begin{figure}[t]
	\centering
	\includegraphics[trim=0 0 0 0,width=0.75\columnwidth]{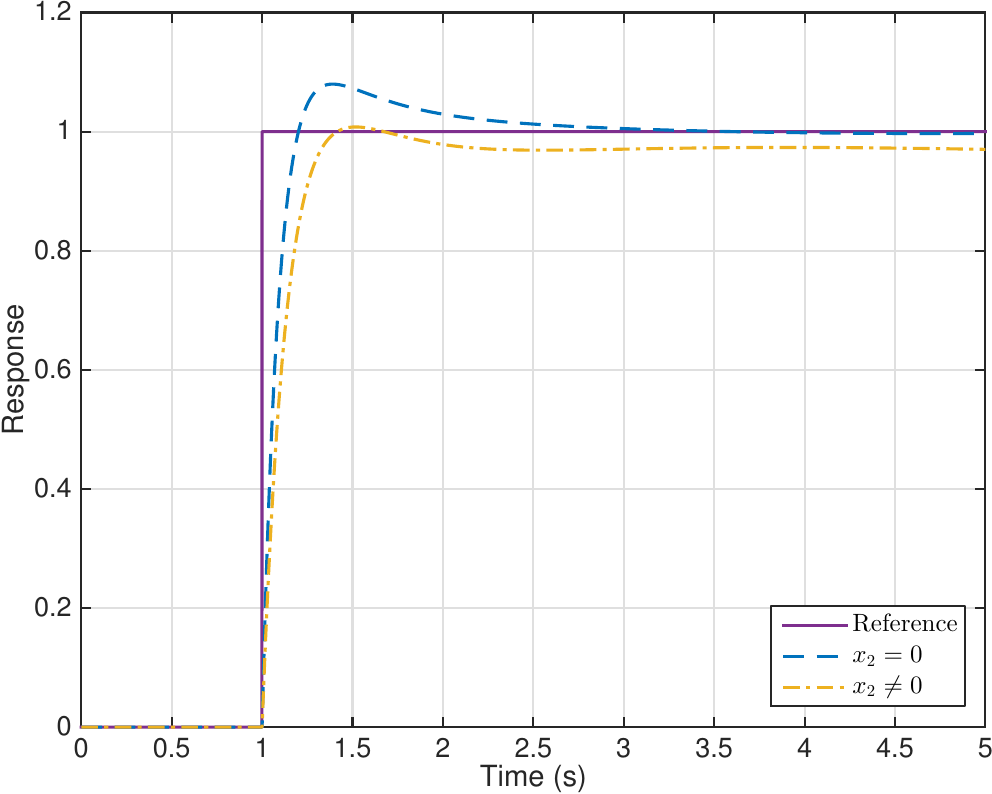}
	\caption{Comparison results of the time response between setting $x_2= 0$ and $x_2\neq 0$ using the proposed robust controller synthesis approach}
	\label{fig:per_x2nonzero}
\end{figure}

Additionally, we perform a test without setting $x_2$ to be zero. With the proposed method, the fixed-order controller parameters are given by $x_1=1.5177$, $x_2=1.5867$, $y_0=13.2818$, $y_1=10.1106$, $y_2=12.6122$, and this controller also stabilizes the closed-loop systems with all prescribed specifications satisfied. In this case, the Bode diagrams of the plant, the controller, and the open-loop system are shown in Fig.~\ref{fig:openloop_x2nonzero}. Also, the Bode diagrams of the sensitivity function and the complementary sensitivity function are depicted in Fig.~\ref{fig:sensitivity_x2nonzero} and Fig.~\ref{fig:csensitivity_x2nonzero}, respectively. Moreover, comparison results of the time response between setting $x_2= 0$ and $x_2\neq 0$ is shown in Fig.~\ref{fig:per_x2nonzero}. When setting $x_2\neq0$, the RMSE  is 0.1103, the maximum overshoot is 0.08\%, and the SSE is 0.0350. Compared with the results when setting $x_2=0$, this test (when setting $x_2\neq 0$) gives worse performance in terms of the RMSE and SSE, but better performance in terms of the overshoot. Note that the user can set any additional requirements on the controller structure, such as $x_2=0$ or $x_2\neq 0$, based on the specific requirements, and the proposed approach will return a feasible solution if one exists.


\section{Conclusion}
This work presented the design and development of a fixed-order controller for SISO systems with interval matrix uncertainties, such that the robust stability and performance under a prescribed finite frequency range is attained. The robust stability condition in terms of SPRness is established, and  the robust performance condition under a finite frequency range is also constructed, with the notion of generalized CL-SPRness and the equivalence between SPRness and SBRness. With this approach, the LMI conditions are formulated to meet the stability criteria and the performance specifications. Rather importantly, it avoids much of the excessive computational efforts and conservatism 
required in various available alternative methods in the existing literature; 
because this approach here presents a new routine 
such that only one set of LMIs is required to to be solved, 
and with restricted frequency ranges in the design specifications taken into account. 
Additionally, a numerical example is provided to validate the theoretical results. 
Suitable and interesting future work would be to incorporate certain time-domain specifications in the formulated problem; for instance, LMIs can also be used for regional pole placement of polytopic systems.

\bibliographystyle{IEEEtran}
\bibliography{IEEEabrv,Reference}

\end{document}